# Passive sub-ambient cooling:
# radiative cooling versus evaporative cooling

Ablimit Aili[a], Xiaobo Yin[a,b], Ronggui Yang[c, *]

[a] Department of Mechanical Engineering, University of Colorado, Boulder, CO 80309, United States
[b] Materials Science and Engineering Program, University of Colorado, Boulder, CO 80309, United States
[c] State Key Laboratory of Coal Combustion, School of Energy and Power Engineering, Huazhong University of Science and Technology, Wuhan, Hubei 430074, China

* Corresponding authors: ronggui@hust.edu.cn

**Abstract**

Day-and-night radiative sky cooling has emerged as a potential alternative to conventional cooling technologies such as refrigeration-based air conditioning and evaporative wet cooling. Both radiative cooling and evaporative cooling can passively achieve sub-ambient cooling without consuming electricity. Although both cooling techniques are subject to impacts from various weather conditions, the extents of the impacts under the same conditions are not well understood. In this work, we experimentally and theoretically study the thermal performances of a passive radiative cooler and a passive evaporative cooler when exposed to a clear night sky. We show that evaporative cooling is better suited for high-temperature and low-humidity weather conditions, with the measured sub-ambient temperatures of the radiative and evaporative coolers being −13.5°C and −15.0°C, respectively, at a low relative humidity of 13% and a high ambient temperature of 26.0°C. On the other hand, radiative cooling is relatively more resilient than evaporative cooling under high-humidity and/or low-temperature weather conditions, with the measured sub-ambient temperatures of the coolers being −11.5°C and −10.5°C, respectively, at a slightly higher relative humidity of 32.0% and a slightly lower ambient temperature of 17.0°C. Depending on water availability and weather conditions, both evaporative cooling and radiative cooling can be adopted as mutually supplemental cooling technologies.

**Nomenclature**

| | |
|---|---|
| $A$ | Sky-facing surface area of a cooler, m$^2$ |
| $c$ | Speed of light, m/s |
| $c_{p,air}$ | Specific heat capacity of air, J/(kg.K) |
| $c_{p,Al}$ | Specific heat capacity of aluminum, J/(kg.K) |
| $c_{p,w}$ | Specific heat capacity of water, J/(kg.K) |
| $D_{vapor}$ | Difffusivity of water vapor in the air, m$^3$/s |
| $h_{air}$ | Air convective heat transfer coefficient, W/(m$^2$.k) |
| $h_{mass}$ | Mass transfer coefficient of evaporation, m/s |
| $h_{fg}$ | Latent heat of vaporization of water, J/kg |
| $h_p$ | Planck's constant, J.s |
| $I_{BB}$ | Spectral irradiance of blackbody, W/(sr.m$^3$) |
| $K_{air}$ | Thermal conductivity of air, W/(m.K) |
| $K_b$ | Boltzmann's constant, J/K |
| $Le$ | Lewis number |
| $m_{Al}$ | Mass of the aluminum plate in a cooler, kg |
| $m_w$ | Initial mass of water in the evaporative cooler, kg |
| $m_{dry}$ | Dry air content of moist air, g/kg |
| $m_{vapor}$ | Water vapor content of moit air, g/kg |
| $P_s$ | Longwave radiation from a cooler surface, W/m$^2$ |
| $P_{atm}$ | Atmospheric longwave radiation absorbed a cooler surface, W/m$^2$ |
| $P_{conv}$ | Convective power density over a cooler, W/m$^2$ |
| $P_{eva}$ | Power density of evaporation from the evaporative cooler, W/m$^2$ |
| $P_{net}$ | Net cooling power of a cooler, W/m$^2$ |
| $PW$ | Precipitable water, mm |
| $p_{amb}$ | Ambient pressure, Pa |
| $p_{dry}$ | Dry air partial pressure (excluding water vapor) , Pa |
| $p_{vapor}$ | Water vapor partial pressure in the air, Pa |
| $R_{dry}$ | Specific gas constant of dry air, J/(kg·K) |
| $R_{vapor}$ | Specific gas constant of water vapor, J/(kg·K) |
| $RH$ | Relative humidity, % |
| $SH$ | Specific humidity, g/kg |
| $T_s$ | Cooler surface temperature, K |
| $T_{amb}$ | Ambient temperature, dry-bulb temperature, K |
| $t$ | Time, s |
| $V_{wind}$ | Wind velocity, m/s |
| $\theta$ | Zenith angle, º |
| $\varphi$ | Azimuth angle, º |
| $\varepsilon_s$ | Spectral emissivity of a surface |
| $\varepsilon_{atm}$ | Spectral emissivity of the atmosphere |
| $\lambda$ | Wavelength of the electromagnetic radiation, m |
| $\rho_{air}$ | Air density (including water vapor in the air), kg/m$^3$ |
| $\rho_{dry}$ | Dry air density (excluding water vapor), kg/m$^3$ |
| $\rho_{v,s}$ | Vapor density at the water surface, kg/m$^3$ |
| $\rho_{v,\infty}$ | Vapor density in the air far from the surface, kg/m$^3$ |
| $\nu$ | Air specific volume, m$^3$/kg |
| χ | Moisture content of air |
| **Abbriviations** | |
| PETG | Polyethylene terephthalate glycol |
| PVDF | Polyvinylidene fluoride |
| Rad | Radiative cocoler |
| Evap | Evaporative cooler |

# 1. Introduction

Developing environment-friendly, energy-efficient, and affordable sub-ambient cooling technologies has become increasingly important as part of efforts to meet the ever-increasing energy demand [1], reduce greenhouse gas emissions [2], combat climate change, tackle water scarcity [3], reduce thermal pollution [4], and address energy and economic poverty [5]. Conventional cooling technologies contribute to the above-listed challenges in many ways, directly or indirectly. For instance, refrigeration-based air conditioning systems may directly leak hydrocarbons that are considered harmful greenhouse gasses or ozone-depleting substances, although refrigerants with reduced environmental damage are being developed [6]. Another drawback of current air conditioning systems is high power consumption and the associated energy bill, which discourages households to turn on air conditioners, even in developed countries [5]. In many developing countries, a large fraction of households have yet to install air conditioning systems because of the installation capital investment and energy costs [7]. Much more severe yet indirect impacts of air conditioning systems come from the fact that fossil-fuel-burning thermal power plants account for most of the electricity generation worldwide [8,9]. Thermal power plants have at least three major impacts on the environment. One is the release of greenhouse gases, mainly $CO_2$ into the atmosphere [10]. Another impact is large water withdrawals and the release of waste heat into water reservoirs, causing thermal pollution [4,11]. Traditional one-through wet-cooled thermal power plants are often associated with water reservoirs' thermal pollutions [11]. The third major impact is evaporative water losses by cooling towers into the atmosphere. On one hand, the water evaporated into the atmosphere is a loss and must be resupplied by a water reservoir, which can otherwise be used for other purposes such as agriculture and domestic applications. On the other hand, the water vapor in the atmosphere amplifies the greenhouse effect [12,13], because it "cloaks" the atmospheric transmittance window, trapping the outgoing thermal radiation from the Earth's surface and increasing the downward radiation from the atmosphere [13,14]. Despite all these challenges associated with conventional cooling technologies, the electricity and water demands for cooling remain high across the globe [16,17].

Considering both cost and efficiency, evaporative cooling has been used in many forms as an alternative to refrigeration-based air conditioning systems or to cool the condensers in thermal power plants [18,19]. The simplest form of evaporative cooling is spraying water on surfaces, which is widely used worldwide to cool ambient and suppress dust on yards, roads, and construction zones. A more sophisticated form of evaporative cooling is swamp coolers used for space cooling. Swamp coolers work best in dry seasons and regions to achieve both humidification and cooling. High humidity, however, severely affects the performance of swamp coolers as the wet-bulb temperature approaches the dry-bulb (ambient) temperature. The most vital application of evaporative cooling is thermal power plant cooling, mostly in the form of wet cooling towers and sometimes cooling ponds. In the US alone, nearly 6 trillion liters out of almost 200 trillion liters of withdrawn water are evaporated into the atmosphere by thermal power plants each year [20,21]. Evaporative cooling in thermal power plants indeed brings several challenges as described previously.

Radiative cooling has been proposed as an alternative cooling technique [22–24]. Without using much electricity or evaporating any water, radiative cooling passively dumps waste heat in the form of infrared thermal emission through the atmospheric window into the deep space, instead of releasing it to the ambient air as conventional cooling systems do [25–27]. Radiative cooling has the potential of achieving deep sub-ambient cooling thanks to the ultra-cold Universe (~2.7K), if little to no solar absorption occurs and parasitic convective loss is minimized [28,29]. With advances in materials science and engineering, highly solar reflective radiative cooling materials have recently been developed and deployed in the form of solid photonic structures [30,31], thin films [32,33], paints [34,35], and even wood [36]. Radiative cooling systems with the capability of sub-ambient cooling of water and cold generation have been demonstrated [37–39].

In water-stressed and hot regions, radiative cooling can be a potential alternative to evaporative cooling [40,41]. Even though evaporative cooling has been widely used and the radiative cooling concept is well accepted and has attracted a lot of research interests, it is not clear how the thermal performances of these two cooling techniques compare under similar environments. They share several similarities and differences. In addition to being able to passively achieve sub-ambient cooling, both are adversely affected by humidity [38,42]. An increase in humidity (strictly speaking, precipitable water) results in a less transparent and more absorptive atmospheric window, thus reducing the net radiative cooling power of a sky-facing emissive surface [38,43,44]. An increase in humidity also results in a reduced water uptake ability of the air, thus diminishing the rate and cooling power of evaporation [42].

Other weather conditions affect radiative cooling and evaporative cooling differently. Convection is considered purely parasitic for achieving sub-ambient cooling by a radiative cooling surface [25]. However, the role of convection in evaporative cooling is two-fold: partially parasitic loss and partially advective gain [42]. An increase in the ambient temperature may reduce the net cooling power of a radiative cooling surface because of increased atmospheric downward radiation and increased convective parasitic loss. On the other hand, a higher ambient temperature causes the atmosphere to uptake more water vapor and thus enhances the evaporative cooling power, even though the convective parasitic loss also increases. These similar yet different effects of weather conditions on radiative cooling and evaporative cooling must be comparatively evaluated when one needs to make an informed choice between the two for practical applications.

In this work, we experimentally and theoretically compare the passive sub-ambient cooling performances of a radiative sky cooler and an evaporative cooler. To ensure the experiment variables are controllable, we carry cooling tests during the night. The lack of solar irradiance, less windy ambient, as well as moderately dynamic ambient temperature and humidity, are helpful to achieve results with better quality. We then use modeling to study the effects of air heat transfer coefficient, ambient temperature, and relative humidity on the sub-ambient cooling performances of the coolers. We also create colormaps of net cooling power as a function of cooler and ambient temperatures under widely different weather conditions: dry weather with low humidity and wet weather with high humidity.

## 2. Modeling

Based on the schematics of the passive radiative and evaporative coolers facing a clear night sky as illustrated in **Fig. 1**, we present the cooling models of the coolers as follows. Both coolers emit longwave thermal radiation into the atmosphere, of which 8-13µm portion can escape into the deep space through transparent atmosphere window. Both coolers are also subject to longwave thermal radiation from the atmosphere and parasitic convective loss. In addition, evaporation of water induced by convection and humidity gradient plays a major cooling role on the evaporative cooler.

### 2.1. Radiative cooling model

For a sky-facing cooler, the power density of upward radiation from the cooler surface is given by,

$$P_s = \int_0^\infty \int_0^{2\pi} \int_0^{\frac{\pi}{2}} \varepsilon_s(\lambda,\theta) I_B(T_s,\lambda) \sin\theta \cos\theta \, d\theta \, d\varphi \, d\lambda, \tag{1}$$

where $I_B$ is the blackbody spectral radiance ($I_B = \frac{2h_p c^2}{\lambda^5} \frac{1}{\exp[h_p c/(\lambda k_B T)] - 1}$), $\lambda$ is the wavelength, and $\varepsilon_s(\lambda,\theta)$ is the hemispherical spectral emissivity of the cooler surface, and $T_s$ is the cooler surface temperature.

The atmospheric downward longwave radiation absorbed by the cooler surface is given by

$$P_{atm} = \int_0^\infty \int_0^{2\pi} \int_0^{\frac{\pi}{2}} \varepsilon_s(\lambda,\theta) \varepsilon_{atm}(\lambda,\theta,PW) I_B(T_{amb},\lambda) \sin\theta \cos\theta \, d\theta \, d\varphi \, d\lambda, \tag{2}$$

where $\varepsilon_{atm}$ is the effective hemispherical spectral emissivity of the atmosphere, and $T_{abm}$ refers to the ambient temperature. Our recent study shows that using zenith-0° atmospheric emissivity, which is the lowest over the hemisphere, and $T_{amb}$, which is the highest in the densest layer of the atmosphere, gives the lowest error when the atmosphere is treated as if it is a solid body with an effective spectral emissivity [15].

The effective atmospheric emissivity, $\varepsilon_{atm}(\lambda,\theta,PW)$, is mainly a function of the atmospheric precipitable water ($PW$), which itself is closely related to both relative humidity and ambient temperature. $PW$ is defined as the thickness of the atmospheric water vapor when condensed into liquid. For a location with a clear sky and a given altitude (~1600 m in this work), $PW$ (in mm) can be estimated by [38,45]

$$PW \approx 2.15 RH \frac{3800 \exp\left(\frac{17.63 T_{amb}}{T_{amb} + 243.04}\right)}{p_{amb}} - 0.82. \tag{3}$$

The atmospheric spectral emissivity can be computed as a function of the precipitable water by using tools such as MODTRAN [15,46,47].

The convective parasitic power density over the cooler surface is simply given by

$$P_{conv} = h_{air}(T_{amb} - T_s), \tag{4}$$

where $h_{air}$ is the air convective heat transfer coefficient. For horizontal rectangular surfaces, it may be expressed as [48–50]

$$h_{air} = a + b V_{wind}, \tag{5}$$

where $V_{wind}$ is the wind velocity, and coefficients as a and b depend on whether there is a windshield or not. Based on the previous studies and a curve fitting method, we estimate the air heat transfer coefficient is $h_{air} = 8.5 + 2.5 V_{wind}$ without a windshield, and $h_{air} = 3.0 + 1.0 V_{wind}$ with a windshield.

Combining equations (1), (2), and (4) along with the radiative cooler parameters and properties, the net cooling power of the passive radiative cooler is then given as

$$P_{net,rad} = P_s(T_{s,Rad}, \varepsilon_{s,Rad}) - P_{atm}(T_{atm}, \varepsilon_{atm}, \varepsilon_{s,Rad}) - P_{conv}(T_{s,Rad}, T_{amb}, h_{air,Rad}). \tag{6}$$

The temperature evolution of the radiative cooler with time is given by

$$AP_{net,Rad} + m_{Al} c_{p,Al} \frac{dT_{s,Rad}}{dt} = 0. \tag{7}$$

where $A$ is the sky-facing surface area of the radiative cooler, $m_{Al}$ is the mass of the cooler surface (mainly the mass of the Aluminum plate in this work).

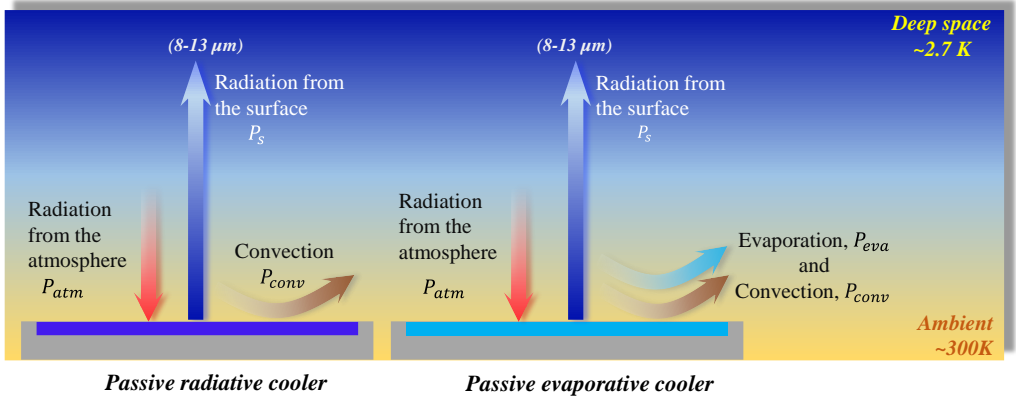

*Fig. 1. Schematic of a passive radiative cooler and a passive evaporative cooler under a clear night sky. Both coolers emit longwave thermal radiation to the atmosphere, of which 8-13μm portion reaches the deep space at 2.7K. Both coolers are subject to longwave thermal radiation from the atmosphere and parasitic convective loss. In addition, evaporation of water induced by convection and humidity gradient plays a major cooling role on the evaporative cooler.*

### 2.2. Evaporative cooling model

For the evaporative cooler, in addition to the radiative and convective terms described above, the evaporation of water is a major cooling mechanism. The power density of evaporation is given by [51]

$$P_{eva} = -\frac{1}{A}\frac{dm_w}{dt} \cdot h_{fg} = h_{mass}(\rho_{v,s} - \rho_{v,\infty})h_{fg}, \tag{8}$$

where $m_w$ is the water mass on the cooler, $h_{mass}$ is the mass transfer coefficient, $\rho_{v,s}$ and $\rho_{v,\infty}$ are respectively the vapor densities at the water surface and in the air far from the surface, and $h_{fg}$ is the latent heat of vaporization of water.

In low mass flux situations, the mass transfer coefficient $h_{mass}$ is analogous to the convective heat transfer $h_{air}$ [51], which is

$$\frac{h_{air}}{h_{mass}} = \rho_{air} c_{p,air} Le^{1-n}, \tag{9}$$

where $\rho_{air}$ is the humid air density, $c_{p,air}$ is the air heat capacity, and $Le$ is the Lewis number, which is the ratio of the thermal and concentration boundary layer thicknesses ($\delta_t/\delta_m = Le^n$). For gases, $Le$ is around unity . Here, the value of the exponent $n$ is ¼ since the surface is horizontal and the mass flux is small [52].

For the air-water system in this work, $Le$ is given by [53]

$$Le = \frac{\rho_{air} c_{p,air} D_{vapor}}{k_{air}}, \tag{11}$$

where $D_{vapor}$ is the mass diffusivity of water vapor in the air, and $k_{air}$ is thermal conductivity of air. $Le$ changes slightly as the air properties change.

The air density is given by its moisture content, $\chi$, and its specific volume based on 1 kg dry air, $v$, as [54]

$$\rho_{air} = \frac{(1+\chi)}{v}. \tag{12}$$

Assuming air and water vapor as ideal gases since their compressibility z factor is close to unity at 1 bar and around 300K [55], the moisture content is given by

$$\chi = \frac{m_{vap}}{m_{dry}} = \frac{R_{air} p_{vap}}{R_{vap} p_{dry}} = 0.622 \frac{p_{vap}}{(p_{amb} - p_{vap})}, \tag{13}$$

where $m_{vap}$ and $m_{dry}$ are the contents of the water vapor and dry air content in the moist air, respectively. $p_{vap}$ and $p_{dry}$ are partial pressures of the water vapor and dry air ($p_{amb} = p_{vap} + p_{dry}$, and $p_{vap} = RH \cdot p_{sat}$), respectively. $R_{vap}$ and $R_{dry}$ are gas constants of water vapor and dry air, respectively.

The specific volume of moist air is given by

$$v = \frac{R_{dry} T_{amb}}{p_{dry}} = \frac{R_{dry} T_{amb}}{p_{amb}} \cdot \frac{p_{dry} + p_{vap}}{p_{dry}} = \frac{1}{\rho_{dry}} (1 + 1.606\chi). \tag{14}$$

Assuming that water vapor in the air is an ideal gas under low pressure [55], the water vapor densities at the cooler surface and in the air can be obtained as

$$\rho_{v,s} = \frac{p_{sat}}{R_{vap} T_{s,Evap}}, \quad \text{and} \quad \rho_{v,\infty} = \frac{RH \cdot p_{sat}}{R_{vap} T_{amb}}. \tag{15}$$

For the evaporative cooler, in addition to the evaporation of water, there are also radiative and convective exchanges between the cooler and sky. Similarly, combining equations (1), (2), (4), and (7) with the evaporative cooler parameters and properties, the net cooling power of the passive evaporative cooler is then given as

$$P_{net,Evap} = P_{eva}(T_{s,Evap}, T_{amb}, RH, h_{mass}) + P_s(T_{s,Evap}, \varepsilon_{eva}) - P_{atm}(T_{atm}, \varepsilon_{atm}, \varepsilon_{s,Evap}) \\ - P_{conv}(T_{s,Evap}, T_{amb}, h_{air,Evap}). \qquad (16)$$

The temperature variation with time of the evaporative cooler is then given by,

$$AP_{net,Evap} + (m_{Al}c_{Al} + m_w c_w)\frac{dT_{s,Evap}}{dt} = 0 \qquad (17)$$

where $c_w$ is the heat capacity of water. The water mass, $m_w$, decreases during the cooling process, as given by Eq.8.

With the transient models for radiative cooling and evaporative cooling presented aboved, we can now evaluate how air heat transfer coefficient, ambient temperature, and relative humidity affect the net cooling power and the sub-ambient temperature of the coolers.

## 2.3.      Effect of air heat transfer coefficient on sub-ambient cooling

Before designing and testing the passive radiative and evaporative coolers, we investigated how the air heat transfer coefficient affects the sub-ambient cooling performances of the coolers under given night-sky conditions of $T_{amb}$ = 25°C and $RH$ = 30%. For sub-ambient cooling, convection is a purely parasitic loss on the radiative cooler, while its role on the evaporative cooler is two-fold: parasitic loss and advective removal of water vapor. Using the spectrum of each cooler's surface material (the coolers are presented in the following Section 3), we calculated the net cooling power of the coolers as a function of their sub-ambient temperatures for different values of $h_{air}$, as shown in **Fig. 2a** and **b.** For the radiative cooler (**Fig. 2a**), the smaller the air heat transfer coefficient, the higher the net cooling power at any sub-ambient temperature. At $P_{net}$ = 0, the sub-ambient temperature of the radiative cooler drops from −4.0°C to −14.0°C as the air heat transfer coefficient $h_{air}$ decreases from 20.0 Wm$^{-2}$K$^{-1}$ to 2.5 Wm$^{-2}$K$^{-1}$. The transition from convective loss to convective gain occurs at zero sub-ambient temperature ($T_s$− $T_{amb}$ = 0) for all $h_{air}$ values.

In comparison, the case for the evaporative cooler is more complicated, as shown in **Fig. 2b**. A low air heat transfer coefficient is beneficial to achieving a sub-ambient cooling temperature as low as possible at $P_{net}$ = 0. For instance, the sub-ambient temperature is −15.0°C and −11.0°C for $h_{air}$ values of 20.0 Wm$^{-2}$K$^{-1}$ and 2.5 Wm$^{-2}$K$^{-1}$, respectively. Another important observation is that all cooling power curves for different $h_{air}$ values intersect at around −10°C sub-ambient temperature. Above this transition point, the larger the air heat transfer coefficient, the lower the sub-ambient temperature at any net cooling power.

To further understand the impact of the air heat transfer coefficient, we plotted the components of net cooling power for $h_{air}$ = 10 Wm$^{-2}$K$^{-1}$ in **Fig. 2c** and **d**. Here, favorable components such as the upward radiation $P_s$ and evaporative power density $P_{eva}$ are plotted as positive terms, whereas undesirable components such as the parasitic convective loss $P_{conv}$ and absorbed downward longwave thermal radiation $P_{atm}$ are plotted as negative terms. On the passive radiative cooler (**Fig. 2c**), the convective loss is more sensitive to the cooler temperature at the given $h_{air}$, ambient conditions, and the shown temperature range. On the evaporative cooler (**Fig. 2d**), the convective loss is initially more sensitive to the cooler temperature below the transition point, above which the evaporative power density becomes more sensitive to the cooler temperature. Interestingly, for a given set of relative humidity and ambient temperature, even if $h_{air}$ value changes, the transition point remains the same, where convective and evaporative power densities have the same rate.

It is important to point out that when there is no sub-ambient cooling ($T_s - T_{amb}$ = 0), the net cooling power densities of the radiative and evaporative coolers are 115 W/m² and at least 190 W/m², respectively, implying that evaporative cooling posses a much higher cooling potential under hot and dry climates.

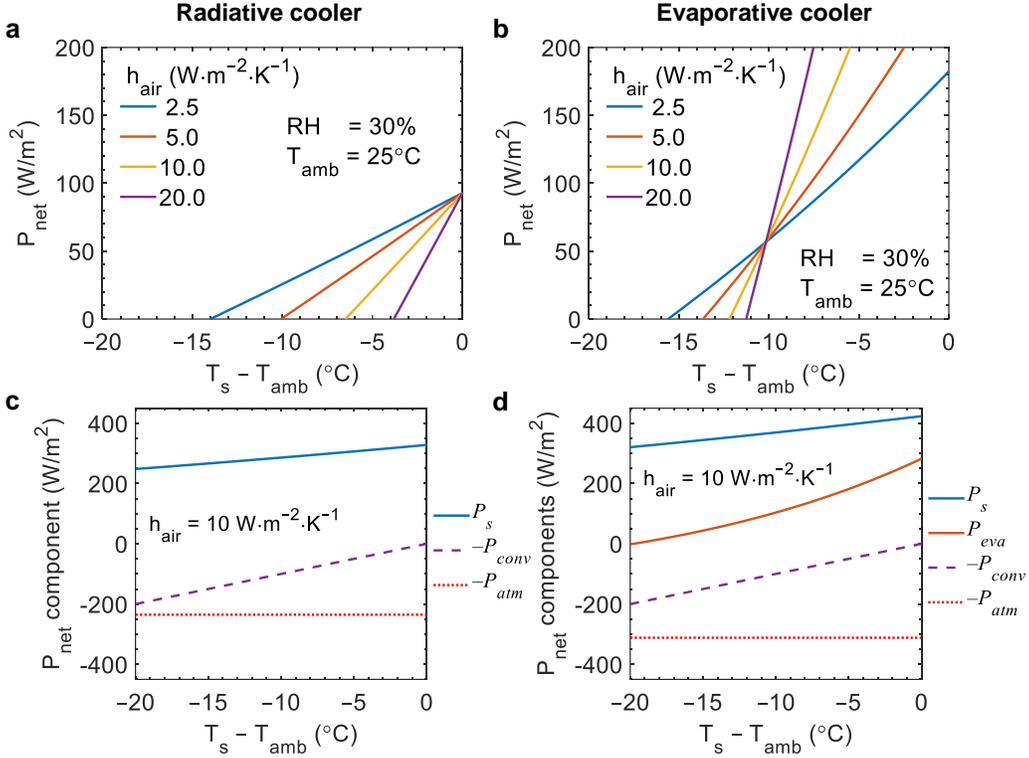

*Fig. 2. Effects of the air heat transfer coefficient ($h_{air}$) on the net cooling power ($P_{net}$) and sub-ambient temperature ($T_s - T_{amb}$) of a passive radiative cooler and a passive evaporative cooler. **(a)** and **(b)**: Net cooling power as a function of the cooler sub-ambient temperature, for the radiative cooler and evaporative cooler, respectively. The ambient conditions are $T_{amb}$ = 25°C and RH = 30%. **(c)** and **(d)**: Components of the net cooling power as functions of the cooler sub-ambient temperature for $h_{air}$ = 10 Wm$^{-2}$K$^{-1}$, for the radiative cooler and evaporative cooler, respectively.*

Based on the analyses above, if the sole purpose in a practical application is to achieve a sub-ambient temperature as low as possible, it is preferable to keep the air heat transfer coefficient small for both coolers, implying that a windshield is beneficial to reduce the convective loss. However, if the purpose is to achieve an acceptable net cooling power with a reasonable sub-ambient temperature above the transition point of the evaporative cooler, the windshield is not beneficial to the evaporative cooler but still is to the radiative cooler. For this reason, a windshield was only added to the radiative cooler during the sub-ambient cooling experiment in Section 3.

## 3. Experiment setup

As shown in **Fig. 3**, a radiative cooling module and an evaporative cooling module with an identical surface area of 0.30×0.30 m$^2$ were prepared to carry out comparative cooling experiments. On the radiative cooler (**Fig. 3a**), a 70-μm thick infrared-emissive film (polyethylene terephthalate glycol, PETG) was laminated on a 1-mm thick aluminum plate. A 10μm thick infrared-transmissive windshield (polyethylene, PE) was added above the cooler to minimize the convective parasitic heat loss. On the evaporative cooler (**Fig. 3b**), a 70-μm thick hydrophobic film (polyvinylidene fluoride, PVDF) was first laminated as a corrosion protection layer on an aluminum plate. A hydrophilic cellulose fabric layer (~100 μm) was placed on top of the film to ensure the uniform spread of water. The initial mass of water at the beginning of the experiment was 600.0g. On both coolers, the back and the four sides were insulated to ensure that heat transfer mainly occur on the sky-facing top surface. Each cooler was equipped with three thermocouples diagonally fixed on the aluminum plate to monitor the cooler temperature during the experiment. Two thermocouples with their tip shielded with solar-reflective aluminum tape were used to monitor the ambient temperature. The wind velocity and relative humidity were measured with a nearby weather station.

**Fig. 3c** shows the spectral emissivity of the radiative cooler surface (polyethylene terephthalate glycol film, PETG) and the evaporative cooler surface (one layer of polyvinylidene fluoride, PVDF, and one layer of cellulose fabric without water). The radiative cooler surface is partially wavelength-selective; that is, it emits thermal radiation mainly in the atmospheric window (8 – 13 μm). The evaporative cooler surface, on the other hand, is nearly black, with a high emissivity at wavelengths beyond 6μm. For sub-ambient cooling, a "selective" radiative cooler is expected to perform better than a "black" radiative cooler [33,38]. However, the addition of water evaporation can enhance the sub-ambient cooling performance of the "black" radiative cooler, i.e., the evaporative cooler.

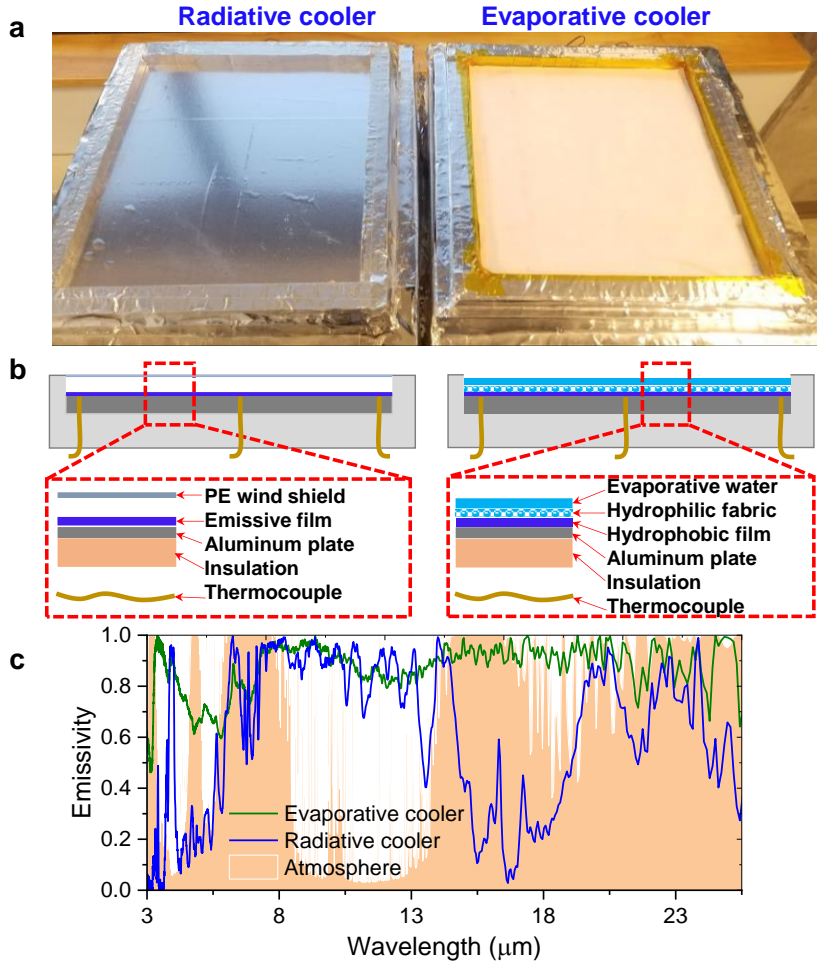

*Fig. 3*. *Experimental setup. (**a**) Optical image of the passive radiative cooler (left) and the passive evaporative cooler (right). The coolers have an identical sky-facing surface area of 0.30×0.30 m². (**b**) Schematics illustrating the design of the coolers. On the radiative cooler, a radiative cooling film (polyethylene terephthalate glycol, PETG, ~70 µm thick) was laminated on a 1-mm thick aluminum plate to ensure the surface temperature uniformity. An infrared transmissive PE windshield was added to minimize the convective loss. On the evaporative cooler, a hydrophobic film (polyvinylidene fluoride, PVDF, ~70 µm thick) was also laminated on an aluminum plate, and then a hydrophilic fabric (cellulose, ~ 100 µm thick) was placed on top of the film to ensure the uniform spreading of water. The initial water mass in the evaporative cooler was 600.0g. On each cooler, three thermocouples were inserted from the back and fixed diagonally on the aluminum plate. (**c**) The spectral infrared emissivity of the radiative cooler surface and the evaporative cooler surface in its dry form, along with the effective atmospheric spectral emissivity for zenith 0° and PW = 5mm. The radiative cooler surface is "partially selective" and the evaporative cooler surface is "nearly black".*

Cooling experiments were carried out over a clear-sky night. There are several advantages in conducting experiments during the nighttime than during the daytime: avoiding complications due to different degrees of solar absorption by the radiative cooler surface and the evaporative cooler surface with water, avoiding fast-changing windy climate during the daytime, reducing frequent fluctuations in the air temperature and relative humidity, and most importantly, achieving controllable and easily comparable cooling performances.

Since a change in one of the weather conditions was often accompanied by changes in other weather conditions during the experiment, we also resort to theoretical approaches in Section 4 to elucidate how radiative cooling and evaporative cooling are affected by changes in a single weather condition while other conditions remain constant.

## 4. Results and discussions
### 4.1. Experimental results

**Fig. 4** presents the measurement results of the nighttime cooling experiment that continued from 8:30 pm to 6:00 am the second day (June 16-17, 2020). The relative humidity and the wind velocity during the experiments are given in **Fig. 4a**. Except for a wind gust, the wind velocity was mostly within 0 ~ 1.0 m/s, implying rather calm weather. Initially, the relative humidity was quite stable within 12 ~ 18%. Upon the arrival of the wind gust, the relative humidity jumped to a range of 30 ~ 35% and remained relatively constant afterward. The ambient temperature (**Fig. 4b**) also evolved at a mostly steady rate, gradually decreasing from a peak of 27°C near the beginning of the experiment to a low of 17°C near the end of the experiment, except for a sudden drop with the arrival of the wind gust.

Mostly stable but occasionally dynamic weather conditions allowed us to observe interesting and subtle changes in the temperatures of the coolers, as presented in **Fig. 4b**. Initially, the coolers were placed indoors at a temperature of 23°C and covered with opaque shields. The coolers were then moved outdoors, and their covers were immediately lifted, triggering rapid temperature drops. There was a lag in the evaporative cooler temperature because of the additional larger thermal mass of water. However, the evaporative cooler temperature fell below the radiative cooler temperature in less than an hour from the start of the experiment, and it remained as such for at least 5 hours. The arrival of the wind gust, a sudden jump in the relative humidity, and a slightly abrupt drop in the ambient temperature all coincided with an inversion in the trend of the cooler temperatures. The radiative cooler temperature now dropped below the evaporative cooler temperature, which is explored further in the following sections.

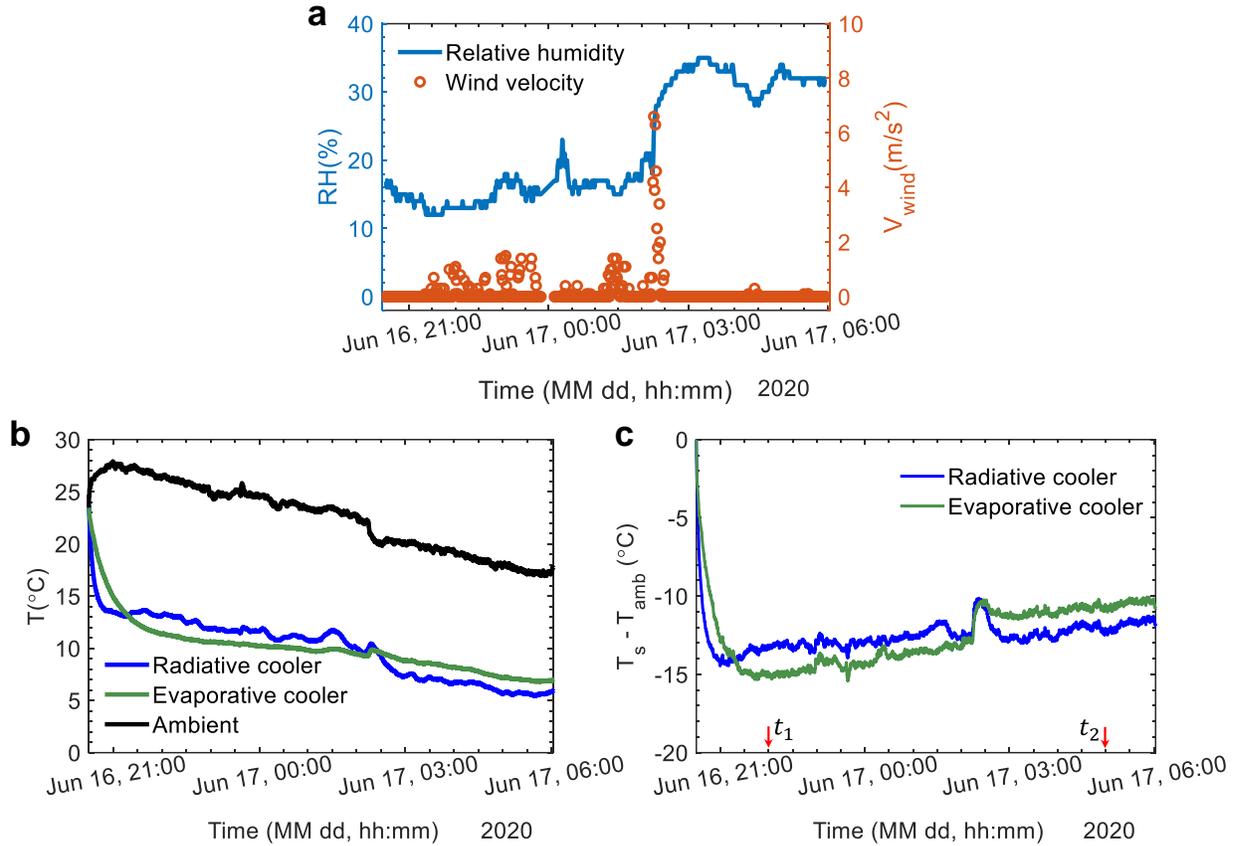

*Fig. 4. Nighttime sub-ambient cooling measurement results. (**a**) Relative humidity and wind velocity. (**b**) Temperatures of the radiative cooler, evaporative cooler, and ambient. (**c**) Sub-ambient temperatures, $T_s - T_{amb}$, of the radiative cooler and the evaporative cooler. The two highlighted time points in **c**, $t_1$ and $t_2$, are selected to present specific sub-ambient temperatures of the coolers under the corresponding weather conditions.*

We also plotted the sub-ambient temperatures ($T_s - T_{amb}$) of the coolers in **Fig. 4c**. Throughout the measurement duration, both coolers' temperatures were well below the ambient temperature. Two timepoints, $t_1$ (10:00 pm) and $t_2$ (05:00 am) indicated by red arrows, are used to specify the sub-ambient temperatures of the coolers. At time $t_1$ and $RH$ = 13.0% and $T_{amb}$ = 26.0°C, the sub-ambient temperatures of the radiative and evaporative coolers were −13.5°C and −15.0°C, respectively, demonstrating the excellent passive cooling performances of both coolers. As the ambient temperature slowly dropped, the sub-ambient temperatures of the coolers gradually increased but were still below −12.0°C, until a flip in trend occurred with the occurrence of the wind gust and abrupt increase in the relative humidity from 12 ~ 18% to 30 ~ 35%. At time $t_2$ and $RH$ = 34.0% $T_{amb}$ = 17.0°C, the sub-ambient temperatures of the radiative cooler and the evaporative cooler were about −12.5°C and −11.0°C, respectively. Although both coolers saw deterioration in their sub-ambient cooling performances as the relative humidity increased from 13.0% at $t_1$ to 34.0% at $t_2$ and as the ambient temperature decreased from 26.0°C to 17.0°C, the degree of deterioration on the evaporative cooling was larger, indicating its cooling power sensitivity to changes in either the ambient temperature or the relative humidity.

In the following sections, we use the theoretical models in Section 2 to further explore the effects of ambient temperature and relative humidity on the performances of the radiative and evaporative coolers under the weather conditions in and beyond the experiment. Using the weather conditions and the coolers' design and property parameters as inputs, we first validate the transient cooling models by comparing the modeled and measured cooler temperatures. As shown in **Fig. 5a** and **b**, the transient cooling models capture reasonably well the temperature dynamics of the radiative and evaporative coolers, respectively.

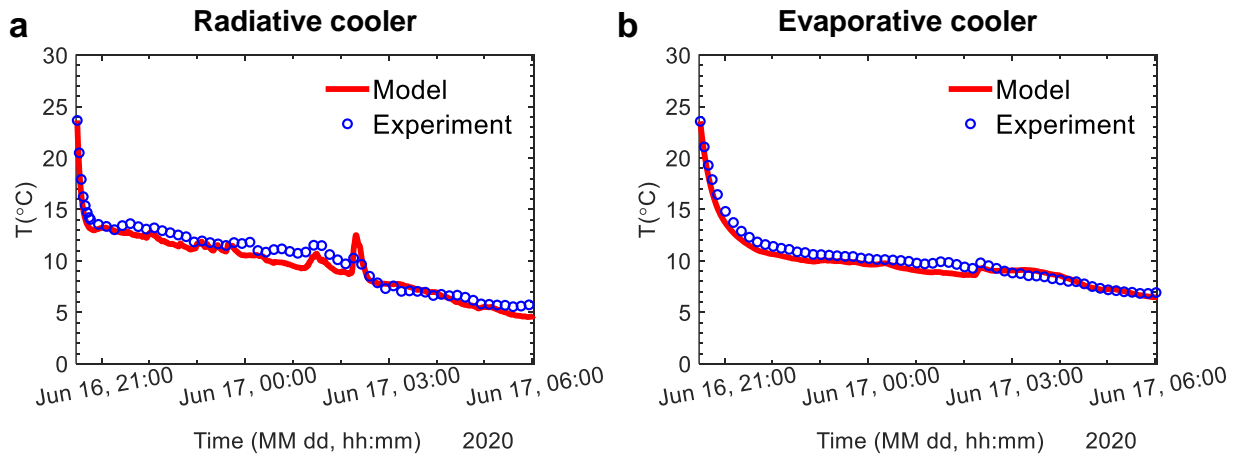

***Fig. 5****. Comparison of the experimental and modeled transient temperatures of coolers over 10 hours of nighttime cooling. (a) Radiative cooler. (b) Evaporative cooler. The experimental data are from **Fig. 4**b.*

### 4.2. Effects of ambient temperature and relative humidity

Different variables of weather conditions, such as the ambient temperature and relative humidity, often change in a partially correlated manner (see **Fig. 4c**), which makes it very difficult to observe directly how the coolers' performances are affected by the changes in only one of such variables. We next use the theoretical models to elucidate how radiative cooling and evaporative cooling are affected by the changes in the ambient temperature or the relative humidity and, consequently, by precipitable water.

We first modeled the equilibrium sub-ambient temperatures of the coolers ($T_s-T_{amb}$) when the relative humidity remains fixed while the ambient temperature is dynamic. **Fig.6a** and **b** show the modeled sub-ambient temperatures of the coolers under two different humidity scenarios, $RH$ = 13.0% and 34%, representing the lower and upper limits of the relative humidity observed during the experiment, respectively. The figures also show the measured sub-ambient temperatures of the coolers for the observed range of ambient temperatures at each specified relative humidity level during the experiment (see **Fig. 4**a). For realistic considerations in modeling, $V_{wind}$ = 1.0 m/s and $V_{wind}$ = 0 are assumed for the two humidity scenarios, respectively (see weather conditions in **Fig. 4**a).

**Fig.6a** and **b** show that the sub-ambient temperatures ($T_s-T_{amb}$) of the radiative and evaporative coolers are affected quite differently by the ambient temperature under the two specified humidity conditions. At a low relative humidity level ($RH$ = 13.%, **Fig.6a**), the sub-ambient temperature of the radiative cooler initially decreases and then slightly increases with the ambient temperature, while the sub-ambient temperature of the evaporative cooler decreases with the ambient temperature. For the radiative cooler, even though relative humidity is fixed, the rise of the ambient temperature results in the increase in the precipitable water at a growing rate (see top x-axes) and thus increasing atmospheric emissivity. The rise of the ambient temperature and its consequent effect on the atmospheric emissivity causes the atmospheric downward radiation to grow at a faster rate (Eq.2) than the upward radiation from the cooler surface as it only depends on the surface temperature (Eq.1). By comparison, for the evaporative cooler, the evaporative cooling power density (Eq.8) increases much faster than other components as the ambient temperature increases, resulting in a larger sub-ambient temperature. At a higher relative humidity level (RH = 34.0%, **Fig.6b**), the radiative cooler performs better than the evaporative cooler, especially as the ambient temperature drops, because of the substantially reduced water uptake ability of air and thus decreased evaporative cooling power.

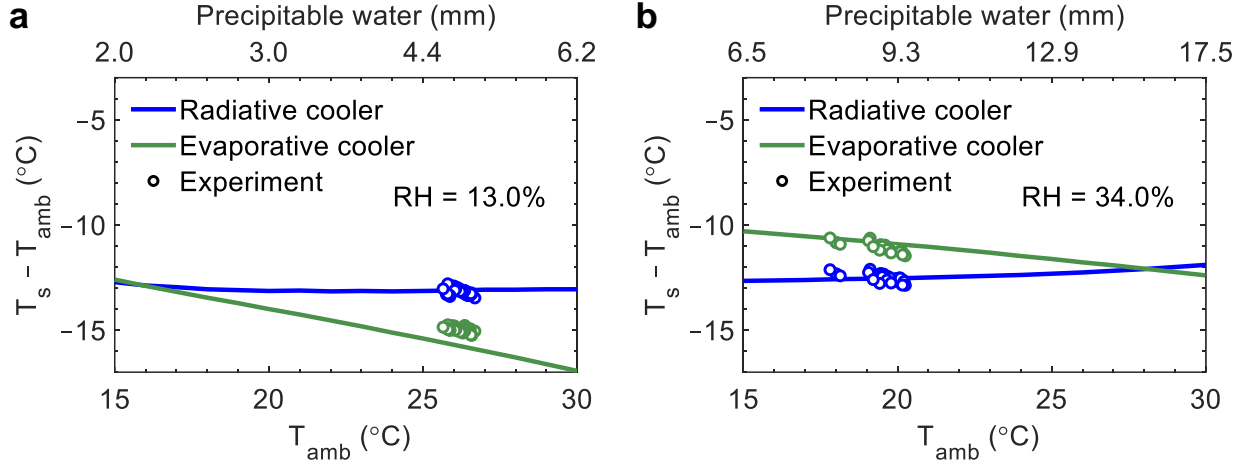

*Fig.6. Sub-ambient temperatures of the radiative and evaporative coolers as a function of the ambient temperature at a fixed relative humidity. **(a)** Cooler performances at RH = 13.0%, near the lower limit of the observed humidity during the experiment. **(b)** Cooler performances at RH = 34.0%, near the upper limit of the observed humidity during the experiment. The experimental results for the observed range of ambient temperatures at the fixed relative humidity are from **Fig. 4c**. The atmospheric precipitable water (PW) estimated from the corresponding ambient temperature and relative humidity is given on the top x-axis. The assumed wind velocity is $V_{wind}$ = 1.0 m/s and 0, for RH = 13.0% and 34.0%, respectively (see **Fig. 4**a).*

**Fig. 7a** and **b** show the effect of the relative humidity on the sub-ambient temperatures of the coolers at two specified ambient temperatures, $T_{amb}$ = 25.0°C and $T_{amb}$ = 17.0°C, respectively, representing the upper and lower limits of the ambient temperature during the experiment. The experimental results for the observed range of relative humidity levels at the specified ambient temperatures from **Fig. 4c** are also presented. Both coolers are adversely affected, yet at different rates, by the increasing relative humidity. At a high ambient temperature, ($T_{amb}$ = 25.0°C, **Fig. 7a**), even though the evaporative cooler mostly performs better than the radiative cooler until the humidity rises above 40%, it sees a faster deterioration in performance than the radiative cooler. At a low ambient temperature, ($T_{amb}$ = 17.0°C, **Fig. 7b**), the radiative cooler overall performs than the evaporative cooler, and its performance worsens at a slower pace than the evaporative cooler. This is because the net cooling power of the radiative cooler is non-linearly and negatively correlated to the relative humidity (Eq.2) through the atmospheric radiation. On the other hand, the net cooling power of the evaporative cooler contains an evaporative term that is linearly and negatively proportional to the relative humidity (Eq. 8) and the atmospheric radiation term that is non-linearly and also negatively correlated to relative humidity (Eq.2). It can be inferred further that the radiative cooler may be a more resilient sub-ambient cooler under high humidity conditions because of its relatively weaker dependence on humidity. On the other hand, the evaporative cooler could perform better under high-temperature and low-humidity conditions.

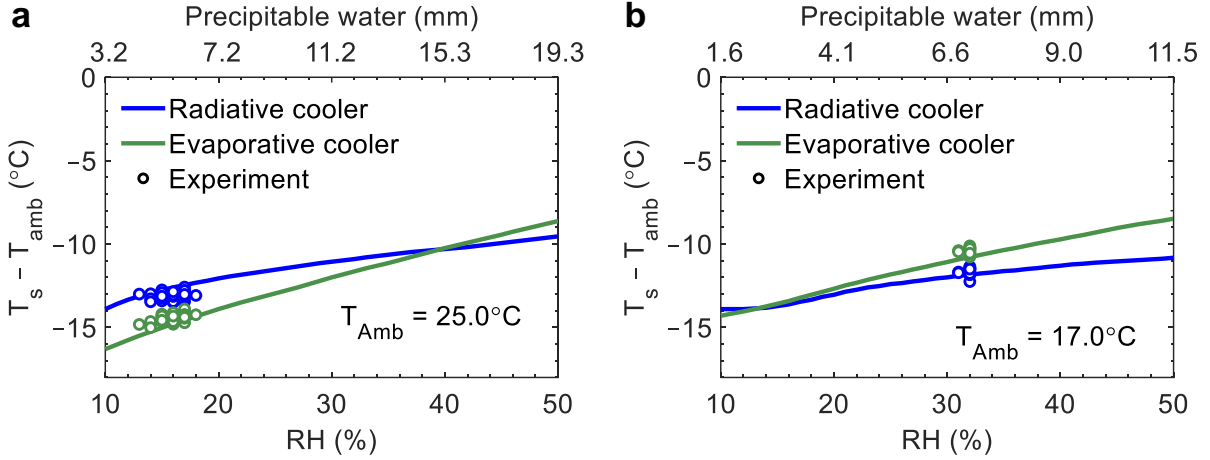

*Fig. 7. Sub-ambient temperatures of the coolers as a function of the relative humidity at a fixed ambient temperature. **(a)** Cooler performances at $T_{amb}$ = 25.°C, near the higher limit of the ambient temperature observed during the experiment. **(b)** Cooler performances at $T_{amb}$ = 17.0°C, near the lower limit of the ambient temperature observed during the experiment. The experimental results for the observed range of relative humidity levels at each fixed ambient temperature are from **Fig. 4c**. The atmospheric precipitable water (PW) estimated from the corresponding relative humidity and ambient temperature is given on the top x-axis. The assumed wind velocity is $V_{wind}$ = 1.0 m/s and 0, for $T_{amb}$ = 25.°C and 17.°C, respectively (see **Fig. 4a**).*

### 4.3. Net cooling power at low and high humidity

To investigate how widely different weather conditions such as very low or very high humidity levels affect the net cooling powers and sub-ambient temperatures of the coolers, we plotted in **Fig. 8 the** net cooling power colormaps as functions of the cooler and ambient temperatures for two humidity values: a low $RH$ = 15% and a high $RH$ = 70%. In this figure, the regimes of sub-ambient cooling are bounded by two dashed curves: the equilibrium sub-ambient temperature curve at $P_{net}$ = 0 and the net cooling power curve at $T_s - T_{amb}$ = 0. The horizontal width of the highlighted regimes represents the sub-ambient temperatures of the coolers. At $RH$ = 15%, both coolers have a large sub-ambient cooling potential at all temperatures shown, although the sub-ambient temperature of the radiative cooler does not change much at high temperatures (**Fig. 8a)**, while the evaporative cooler sees a noticeable increase (**Fig. 8b**). At *RH* = 70%, however, the sub-ambient cooling potential of both coolers diminishes at all temperatures (**Fig. 8c** and **d**), which is especially significant on the evaporative cooler. The sub-ambient temperature of the radiative cooler also becomes significantly small at high temperatures and this specified high humidity.

The cooling power colormaps and the sizes of the sub-ambient cooling regimes indicate that evaporative cooling is better suited for high-temperature and low-humidity weather conditions, although radiative cooling also performs well under such conditions. On the other hand, radiative cooling is positioned better than evaporative cooling to work under high-humidity and low-temperature weather conditions, where cold generation and storage for reuse may be a more practical approach. Unfortunately, high-humidity and high-temperature weather conditions are detrimental to the performances of both coolers.

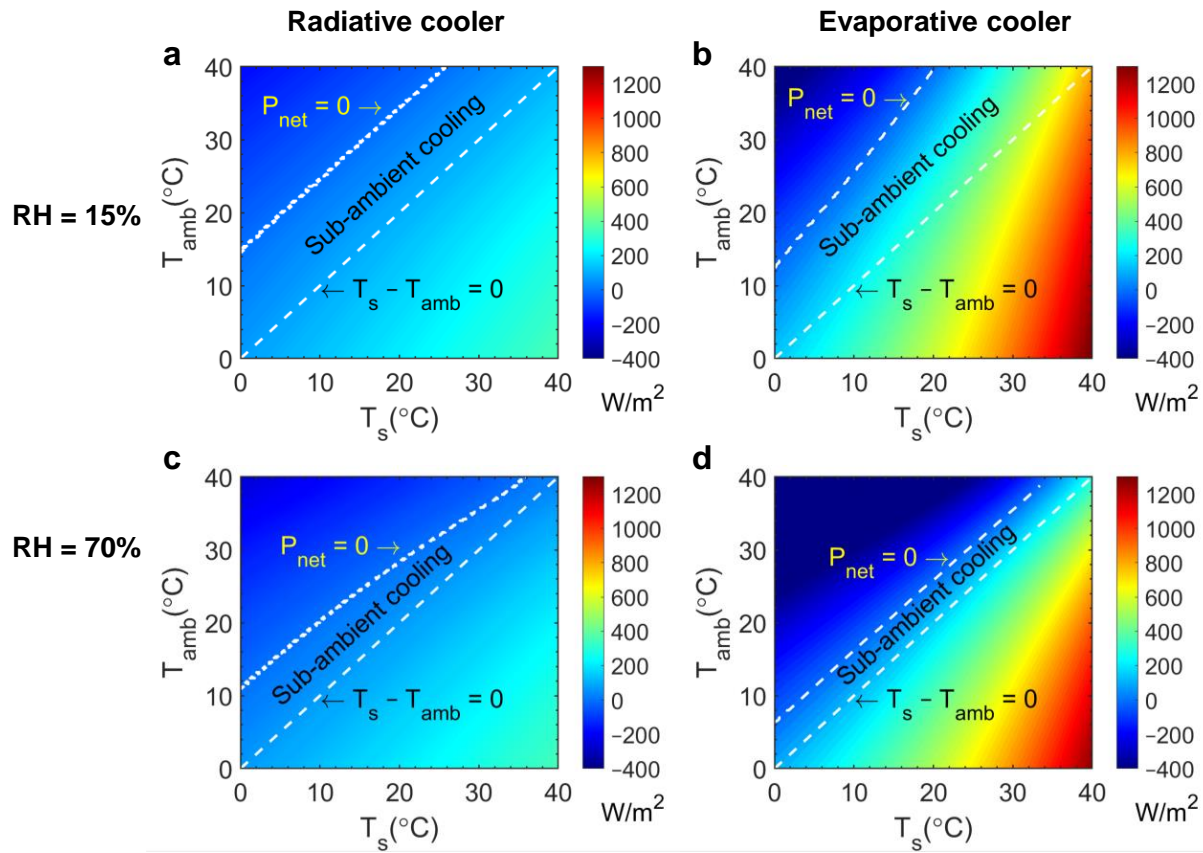

*Fig. 8. Net cooling power colormaps as functions of the cooler surface temperature and the ambient temperature. (**a**) Radiative cooler and (**b**) evaporative cooler. The two figures at the top are for a low-humidity condition (RH = 15.0%), and the two figures at the bottom are for a high-humidity condition (RH = 70%). The regions highlighted by the two dashed lines represent passively achievable sub-cooling regimes defined by*

$T_s - T_{amb} = 0$ and $P_{net\ cool} = 0$. The horizontal width of the sub-ambient cooling regimes represents the sub-ambient temperatures of the coolers.*

## 5. Conclusions

In this work, we experimentally and theoretically investigated the sub-ambient cooling performances of radiative cooling and evaporative cooling under the same weather conditions. With sky-facing radiative and evaporative coolers, we experimentally observed that the sub-ambient temperatures ($T_s - T_{amb}$) of the coolers were −13.5°C and −15.0°C, respectively, at a relative humidity ($RH$) of 13.0% and an ambient temperature ($T_{amb}$) of 26°C. This shows that the evaporative cooler performs better than the radiative cooler at low-humidity and high-temperature weather conditions. At a higher relative humidity of 34.0% and a lower ambient temperature of 17°C, the observed sub-ambient temperatures of the radiative and evaporative coolers were −12.5°C and −11.0°C, respectively. This indicates the radiative cooler is better positioned to work under low-temperature or high-humidity conditions.

Theoretical models were developed and validated to elucidate the impacts of convection, ambient temperature, and relative humidity on the sub-ambient cooling performances of the coolers. We then created net cooling power colormaps as functions of the cooler and ambient temperatures for two humidity conditions, a low $RH$ = 15% and a high $RH$ = 70%. The cooling power colormaps and the sizes of the sub-ambient cooling regimes not only further confirm our experimental observations but also provide more information regarding applications under conditions beyond that observed in this work.

Our study validates and compares the passive sub-ambient cooling potential of radiative cooling and evaporative cooling. In regions with high humidity or limited water resources, radiative cooling can be a potential alternative to evaporative cooling. Depending on water availability and weather conditions, both evaporative cooling and radiative cooling can be adopted as mutually supplemental cooling technologies.